\documentclass[twocolumn,showpacs,prl,aps]{revtex4}

\usepackage{graphicx,color}
\usepackage{dcolumn}
\usepackage{bm}

\begin{document}

\title{
Mesoscopic Hall effect driven by chiral spin order
}

\author{Jun-ichiro Ohe$^{1}$}

\author{Tomi Ohtsuki$^{2}$}

\author{Bernhard Kramer$^{3}$}

\affiliation{%
$^{1}$I. Institut f\"{u}r Theoretische Physik, Universtit\"{a}t Hamburg,
Jungiusstrasse 9, 20355 Hamburg, Germany \\
$^{2}$Department of Physics, Sophia University,
Kioi-cho 7-1, Chiyoda-ku, Tokyo 102-8554, Japan \\
$^{3}$School of Engineering and Science, 
International University Bremen, Campus Ring 1,
28757 Bremen, Germany}

\date{\today}

\begin{abstract}
A Hall effect due to 
spin chirality in mesoscopic systems is predicted.
We consider a 4-terminal Hall system including local spins
with geometry of a vortex domain wall, where strong spin chirality
appears near the center of vortex.
The Fermi energy of the conduction electrons is assumed to be comparable
to the exchange coupling energy
where the adiabatic approximation
ceases to be valid.
Our results show
a Hall effect where a voltage drop and a spin
current arise in the transverse direction.
The similarity between this Hall effect and
the conventional spin Hall effect in systems with spin-orbit interaction is pointed out.

\end{abstract}


\maketitle

Recent research on the anomalous Hall effect has shown
that the spin chirality of a local spin system induces a
Hall conductance via exchange coupling
\cite{Taguchi,Onoda,tatara:113316,PhysRevB.62.R6065}.
The anomalous Hall effect can be seen in ferromagnetic metallic systems
where the time reversal symmetry (TRS) is broken.
When TRS is preserved, the spin current in a transverse
direction is driven by a longitudinal voltage drop.
Such a spin current, the so-called spin Hall current
\cite{Murakami,Sinova,Kato},
can be seen in
 semiconductor systems with a spin-orbit interaction.
Both anomalous and spin Hall effects were originally expected
in bulk systems where the
gauge field related to monopoles in momentum space
plays a crucial role.
The spin Hall effect can also be seen in
mesoscopic 2-dimensional samples with
 Rashba spin-orbit interaction \cite{Sinova}.
Numerical calculations, using both the Kubo and Landauer-Buttiker formulae
predict the spin Hall effect.
Note that the Landauer-Buttiker formula \cite{Landauer,Buttiker}
does not explicitly assume
a local electric field inside the sample \cite{Dattatxt},
while such a field
seems to be essential for explaining the conventional spin Hall effect
\cite{Murakami}.

Recently, it has been shown that a Hall conductance is expected in
mesoscopic systems such as dilute magnetic semiconductors
 with artificial magnetic structures \cite{Bruno,bruno:247202}.
The spin of the conduction electron couples to the
local magnetic moment via an exchange interaction.
The Hall conductance is determined in such a
well ordered magnetic system
by using a local gauge transformation and the adiabatic approximation
in which only a majority spin component is considered.
This approximation changes the symmetry of the system
from SU(2) to U(1).
However, the minority spin of conduction electrons cannot be
neglected when the exchange coupling energy is comparable to the 
Fermi energy as in magnetic semiconductors \cite{furdyna:R29}.

\begin{figure}[bbb]
\begin{center}
\includegraphics[scale=0.35,angle=0]{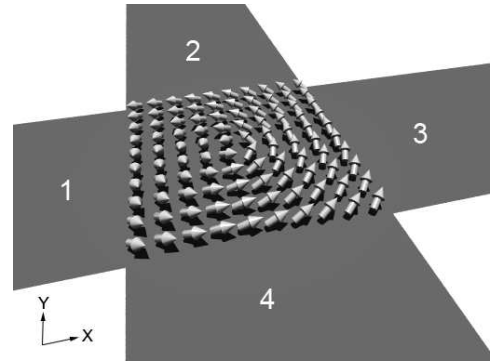}
\caption{\label{ohefig:1}
  Schematic of proposed 4-terminal system, including
chiral magnetic structure.  Labels
1-4 indicate leads that are free from randomness,
local magnets, or spin-orbit interaction.}
\end{center}
\end{figure}

In this letter, we show that mesoscopic systems with internal
chiral magnetic order exhibit Hall effects
in such a way that 
both the charge and the spin Hall effects occur simultaneously.
We consider a 2-dimensional electron system that interacts with
local spins via exchange coupling.
The local spins have a vortex structure with a finite
out-of-plane component that determines the spin chirality.
We assume that the Fermi energy of the conduction electron
is comparable to the exchange coupling energy as in some
magnetic semiconductors.
In such energy region, the adiabatic approximation and
the U(1) mean field
theory \cite{PhysRevB.62.R6065} that explains the anomalous Hall effect
cannot be applied.
We calculate
the spin-resolved  Hall conductance numerically
 by using the recursive Green
function method \cite{Ando91,Yamamoto}.
Our numerical results show that a Hall voltage is induced
when the system has spin chirality.
Furthermore, a spin Hall current can be observed
even if the system does not have a spin chirality.
This spin Hall current does not 
require an electric field inside the system unlike
 conventional spin Hall effects in bulk systems with
spin-orbit interaction
\cite{Murakami}.
We mention that the system considered here
is related to a 2-dimensional spin-orbit 
system for which the spin Hall effect has been reported.
We show that the previously reported spin Hall effects
\cite{Sheng_PRL94_016602,Nikolic_PRB72_075361}
obtained from the Landauer-Buttiker formula are similar to the Hall effect
presented here.
We investigate the coupling constant dependence of the spin Hall
conductances for these systems.
Both of the spin Hall conductances oscillate when increasing the
exchange or spin--orbit coupling strength and show linear dependences
in the weak coupling regime.

\begin{figure}[ttt]
\begin{center}
\includegraphics[scale=0.45,angle=0]{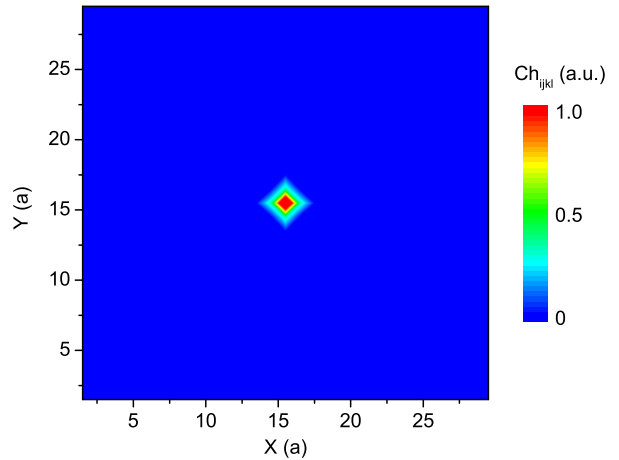}
\caption{\label{ohefig:2}
Spatial configuration of the spin chirality defined
by Eq.~(\ref{oheeq:ch}).
System size is set to $30a \times 30a$.
Parameters are $S=1$ and $\theta=0.3 \pi$.
Strong spin chirality is seen near the center.
}
\end{center}
\end{figure}

We consider a 2-dimensional electron system with exchange
interaction \cite{Ohe03},
\begin{eqnarray}
H=-t\sum_{<i,j>,\sigma,\sigma'}c^+_{i\sigma}c_{j\sigma'} 
-J\sum_{i,\sigma,\sigma'}c_{i\sigma}^+\boldsymbol{\sigma}
c_{i\sigma'}
\cdot {\bf S}(x,y),
\label{oheeq:H}
\end{eqnarray}
with the
nearest-neighbor hopping parameter $t={\hbar^{2}}/{2m^* a^{2}}$
($m^*$ effective mass, $a$ lattice parameter).
Operator $c^+_{i\sigma} (c_{i\sigma})$ creates
(annihilates) an electron of spin $\sigma$ at lattice site $i$,
$\boldsymbol{\sigma}$'s are the Pauli matrices,
and $J(>0)$ is the exchange coupling constant.
The local spin ${\bf S}(x,y)$ has the geometry of a vortex in the $x$-$y$
plane in addition to the uniform $S_z$-component
\begin{eqnarray}
{\bf S}(x,y)=
S(\cos\phi(x,y)\sin\theta,\sin\phi(x,y)\sin\theta,\cos\theta).
\end{eqnarray}
Here, $S$ is the modulus of the local spin,
$\phi(x,y)=-\tan^{-1}(y/x)$, such that
the center of the vortex is located at the origin.
We assume that the dynamics of the local spins is much slower
than that of the conduction electrons,
and treat the local spins as static.
A schematic of the system is shown in Fig.~\ref{ohefig:1}
in which we assume 4-terminal geometry.
The leads are labeled as $1$-$4$ and the $+x(y)$-direction is set to 
the direction from the lead $1(4)$ to $3(2)$.
We define the chirality of the local spin system,
\begin{eqnarray}
{\rm Ch}_{ijkl}\equiv  E_{ijk}+E_{ikl},
\label{oheeq:ch}
\end{eqnarray}
where $E_{ijk}={\bf S}_i \cdot ({\bf S}_j \times {\bf S}_k)$ for a plaquette of
a square lattice labeled as $(i,j,k,l)$ counter--clockwise.
Fig.~\ref{ohefig:2} shows the distribution of the spin chirality for
$\theta=0.3 \pi$.
The system shows large spin chirality near the center of the vortex.
The value of the spin chirality changes the sign when the
sign of the $S_z$-component changes.
Obviously, the local spin system does not have a chirality,
 ${\rm Ch}_{ijkl}= 0$, for
$\theta = 0, \pi/2, \pi$.
As in the case of the bulk anomalous Hall effect due to spin chirality
\cite{PhysRevB.62.R6065}, we expect that the
Hall effect can be obtained except for these values of $\theta$.

We calculate the spin-resolved transmission amplitudes
by using the recursive
Green function method \cite{Ando91,Yamamoto}.
By employing the Landauer-Buttiker formula,
we assume that the net current of the leads 2 and 4 is zero.
The current of lead $I_l$ is
$I_l=\sum_{\sigma}(N_l-R_{l\sigma,l\sigma})\mu_l-\sum_{l\neq l'\sigma\sigma'}
T_{l\sigma,l'\sigma'}\mu_{l'}$,
where $N_l$ is the number of propagating channels per spin for the lead $l$,
$T_{l\sigma,l'\sigma'}(R_{l\sigma,l'\sigma'})$
is the transmission (reflection) amplitude
from the $\sigma'$-spin channel of lead $l'$ to the $\sigma$-spin channel of the lead $l$, and
$\mu_l$ is the chemical potential of the reservoir attached to the lead $l$.
The Hall conductance is defined as
$G_{\rm H}= -r_{yx}/(r_{xx}^2+r_{yx}^2)$,
where $r_{yx}=(\mu_2-\mu_4)/I_1$ and $r_{xx}=(\mu_1-\mu_3)/I_1$
are the Hall resistance and the
resistance, respectively.
The spin Hall conductance is defined as
$G_{{\rm sH}}^{\nu}=(I_{2+}^{\nu}-I_{2-}^{\nu})/(\mu_1-\mu_3)$,
where $I_{2\pm}^{\nu}$ is the current in the lead $2$ with a polarization
in the $\pm \nu\,(\nu=X,Y,Z)$ direction.

\begin{figure}[ttt]
\begin{center}
\includegraphics[scale=0.42,angle=0]{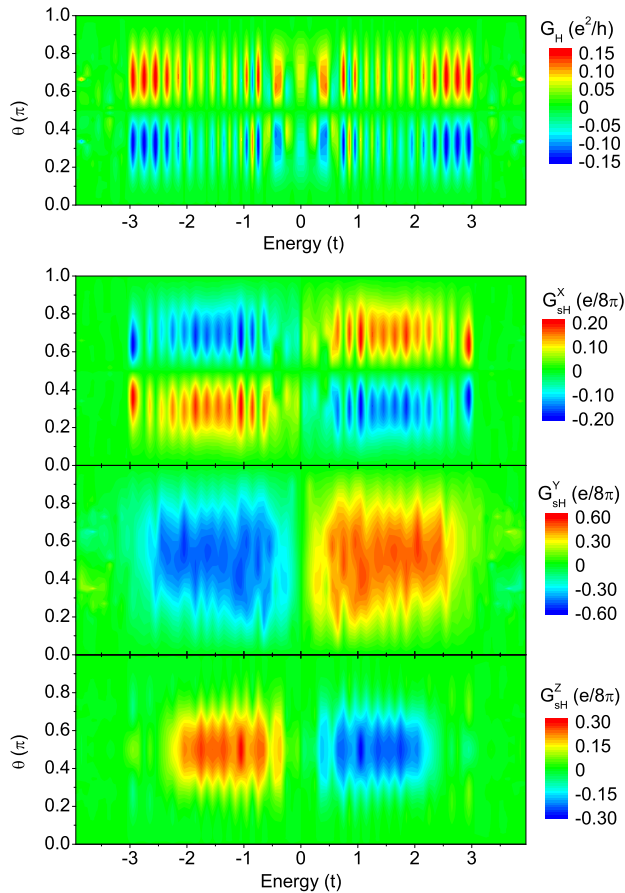}
\caption{\label{ohefig:3}
Hall conductance and spin Hall conductance for each polarization:
Exchange coupling constant is $JS=t$, and
the system size $30a \times 30a$.
Hall conductance and $x$-component of the
spin Hall conductance disappear at $\theta=0, \pi/2, \pi$,
where the spin chirality vanishes.
}
\end{center}
\end{figure}

The uppermost panel of Fig.~\ref{ohefig:3} shows the Hall conductance as a function of the
energy of the conduction electrons and the angle of the local spin $\theta$
for $JS=1.0\,t$ in a system of size $30a\times 30a$.
The Hall conductance is non zero for $\theta \neq 0, \pi/2, \pi$.
Its sign changes when the sign of
$S_z$ changes.
The amplitude of the Hall conductance oscillates with 
the energy.
This is because
the Hall effect is proportional to the momentum of the $x$-direction, hence
it decreases when the Fermi energy
is close to the energy where new propagating channels open.
We note that the (charge) Hall effect also induces
spin current density to be polarized parallel to the 
local spins near the interface between the lead 2 (or 4) and the sample.

If the Fermi energy is comparable to the exchange
coupling energy,
the adiabatic approximation that neglects the minority spin
components cannot be applied,
and we expect a spin current with a polarization that is not parallel
to the local spins.
To confirm this, we plot the spin Hall conductance for each polarization direction
in the lower 3 panels of Fig.~\ref{ohefig:3}.
$G_{\rm sH}^{Y,Z}$ vanish at $\theta=0, \pi$ while
$G_{\rm sH}^X$ vanishes at $\theta=0, \pi/2, \pi$.
At $\theta=\pi/2$,
the local spins near the interface between
the sample and the
lead 2 almost direct in the $x$-direction, and
the suppression of the Hall conductance results in a 
suppression of $G_{\rm sH}^X$.
The non-vanishing $Y$- and $Z$- components of the
spin Hall conductance at $\theta=\pi/2$
do not induce a voltage drop in the
transverse direction like the spin Hall effect
predicted in the spin-orbit system \cite{Sheng_PRL94_016602,Nikolic_PRB72_075361}.
The direction of polarization rotates while electrons propagate in the
sample due to the precession induced by the exchange coupling.
This precession is an important feature of the mesoscopic spin Hall effect
that is also obtained in a 2-dimensional electron system with
spin-orbit interaction.
In contrast, only $Z$-component of the spin current is expected
in bulk spin-orbit systems.

\begin{figure}[bbb]
\begin{center}
\includegraphics[scale=0.4,angle=0]{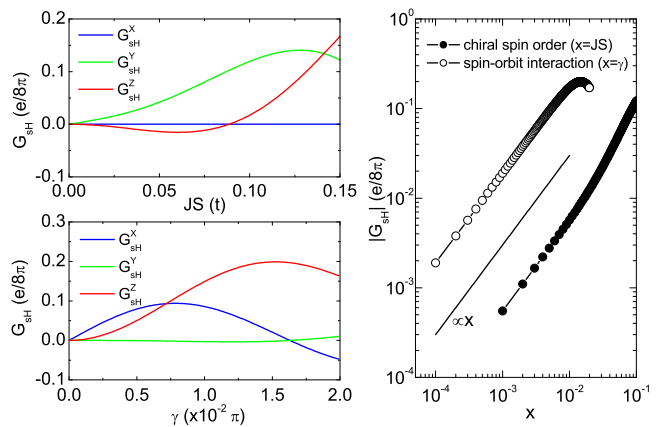}
\caption{\label{ohefig:4}
Coupling parameter dependences of the spin Hall conductances
of the chiral spin order system 
with $\theta=\pi/2$ (left-upper panel)
and of
the Rashba spin-orbit interaction system (left-lower panel).
The right panel shows the log-log plot of absolute values
of spin Hall conductances
in a weak coupling regime. Both spin Hall conductances show a
linear dependence on the coupling parameters $JS$ and $\gamma$.
}
\end{center}
\end{figure}

To make contact with the 2-dimensional system with spin-orbit interaction,
we consider the Rashba spin-orbit interaction represented as 
\begin{eqnarray}
H=-t \sum_{<i,j>,\sigma,\sigma'} V_{i\sigma,j\sigma'}
    c^{\dagger}_{i\sigma}c_{j\sigma'}
\label{oheeq:Hso}
\end{eqnarray}
with
\begin{eqnarray}
V_{i+\hat{\bf x},i}=
\left(
\begin{array}{cc}
\cos\gamma & \sin\gamma \\
-\sin\gamma & \cos\gamma
\end{array}
\right),
\end{eqnarray}
and
\begin{eqnarray}
V_{i+\hat{{\bf y}},i}=
\left(
\begin{array}{cc}
\cos\gamma & -{\mathrm i}\sin\gamma \\
-{\mathrm i}\sin\gamma & \cos\gamma
\end{array}
\right),
\end{eqnarray}
where $\gamma$ is
the coupling strength of the Rashba spin-orbit interaction
\cite{Rashba,Ando,Nitta}.
The parameters $JS$ and $\gamma$ can be regarded as a gauge field strength
\cite{Aleiner_PRL87_256801,Aleiner_PRL87_256801_Erratum}.
Fig.~\ref{ohefig:4} shows the dependence on the coupling strength of
the spin Hall conductances both for the chiral
spin system and for the spin-orbit system.
The spin Hall conductances oscillate with the coupling constant
in both cases.
Indeed, the component of the spin Hall conductance in the spin-orbit system
shows spin precession by changing the length of the lead
where the spin-obit interaction is present \cite{Oheun}.
This oscillation cannot be obtained in bulk system
with a spin-orbit interaction and a local electric field,
where a monopole in momentum space description is possible.
In this sense, the spin Hall current obtained in the present paper
 should be distinguished from the bulk spin Hall effect described
by the Kubo formula.
We also show the absolute value of the spin Hall conductance
$|G_{\rm sH}|
=\sqrt{{G_{\rm sH}^{X}}^2+{G_{\rm sH}^{Y}}^2+{G_{\rm sH}^{Z}}^2}$
in the weak coupling regime.
Here, both spin Hall conductances show linear dependence
on the coupling constant in weak coupling regime.

For measuring the present Hall effect in actual systems,
an experimental setup using ferromagnetic semiconductors,
such as (Ga,Mn)As, can be used.
The proposed vortex spin configuration \cite{Shinjo} can be obtained
in dilute magnetic semiconductors with
low Curie temperatures \cite{PhysRevB.64.241201,Nazmul}.
Because of the small saturation magnetization ($\approx 0.01$\, T)
of magnetic semiconductors, the coupling
energy between the conduction spin and local magnetic moment
should be comparable to the Fermi energy ($\approx$\,meV) \cite{furdyna:R29}.
For our calculations, by setting the tight binding parameter
$a=10$\,nm and $m^*=0.05$\,$m_e$, 
the corresponding exchange energy becomes $J\approx 6.9$\,meV.
$\theta$ should be adjusted approximately to $\pi/2$ to minimize
heating effect that destroys the spin order.

In conclusion, we have investigated a new mesoscopic Hall effect
driven by a local spin system with spin chirality, which
might be experimentally detected in 2DES embedded in
ferromagnetic semiconductors.
The local spin system is assumed to have the geometry of a vortex with
a chirality at the center.
We have predicted a Hall effect,
which induces both a charge and a spin Hall conductance.
Our numerical results based on the Landauer-Buttiker formula
and the recursive Green function technique
show that a voltage drop is obtained in the presence of spin chirality.
No uniform electric field is required inside the sample.
We have pointed out that the present Hall effect is
related to the spin Hall effect obtained for a 2-dimensional 
spin-orbit system,
but should be distinguished
from the usual bulk spin Hall effect driven by
monopoles in momentum space described by the
Kubo formula.

The effect of randomness on the charge and the spin Hall effect, which is
a problem left for the future, is also interesting.
For $\theta=\pi/2$, the Hamiltonian becomes 
a real matrix via
the unitary transformation $U=\exp(-\mathrm{i}\pi\sigma_x/4)$, and
the system is mapped to an orthogonal system with a spin Hall current.

\begin{acknowledgments}
The authors are grateful to M. Yamamoto, S. Kettemann and Y. Avishai
for valuable discussions.
This work has been supported by the
Deutsche Forschungsgemeinschaft via SFBs 508 and 668 of the
Universit\"at Hamburg, and by the European Union via the
Marie-Curie-Network MCRTN-CT2003-504574.
\end{acknowledgments}


\begin{thebibliography}{99}
\bibitem{Taguchi} 
Y. Taguchi, Y. Oohara, H. Yoshizawa, N. Nagaosa and Y. Tokura,
Science {\bf 291,} 2537 (2001).

\bibitem{Onoda} 
S. Onoda and N. Nagaosa,
Phys. Rev. Lett. {\bf 90,} 196602 (2003).

\bibitem{tatara:113316}
G. Tatara and H. Kohno,
Phys. Rev. B {\bf 67,} 113316 (2003).

\bibitem{PhysRevB.62.R6065}
K. Ohgushi, S. Murakami and N. Nagaosa,
Phys. Rev. B {\bf 62,} 6065(R) (2000),

\bibitem{Murakami}
S. Murakami, N. Nagaosa and S. -C. Zhang,
Science {\bf 301} 1348, (2003).

\bibitem{Sinova}
J. Sinova, D. Culcer, Q. Niu, N. A. Sinitsyn, T. Jungwirth and and A. H. MacDonald,
Phys. Rev. Lett. {\bf 92,} 126603 (2004).

\bibitem{Kato}
Y. K. Kato, R. C. Myers, A. C. Gossard and D. D. Awschalom,
Science {\bf 306} 1910, (2004).

\bibitem{Landauer}
R. Landauer, IBM J. Res. Dev. {\bf 1,} 223 (1957).

\bibitem{Buttiker}
M. B\"uttiker, Phys. Rev. Lett. {\bf 57,} 1761 (1986).

\bibitem{Dattatxt}
S. Datta, {\it Electronic Transport in Mesoscopic Systems}
(Cambridge University Press, Cambridge, 1995).

\bibitem{Bruno}
P. Bruno, V. K. Dugaev and M. Taillefumier,
Phys. Rev. Lett. {\bf 93,} 096806 (2004).

\bibitem{bruno:247202}
P. Bruno,
Phys. Rev. Lett. {\bf 93,} 247202 (2004).

\bibitem{furdyna:R29}
J. K. Furdyna,
J. Appl. Phys. {\bf 64,} R29 (1988).

\bibitem{Ando91}
T. Ando,
Phys. Rev. B {\bf 44,} 8017 (1991).

\bibitem{Yamamoto}
M. Yamamoto, T. Ohtsuki and B. Kramer,
Phys. Rev. B {\bf 72,} 115321 (2005).

\bibitem{Sheng_PRL94_016602}
L. Sheng, D. N. Sheng and C. S. Ting,
Phys. Rev. Lett. {\bf 94,} 016602 (2005).

\bibitem{Nikolic_PRB72_075361}
B. K. Nikoli\'c, L. P. Z\^arbo and S. Souma,
Phys. Rev. B {\bf 72,} 075361 (2005),

\bibitem{Ohe03} J. -I. Ohe, M. Yamamoto and T. Ohtsuki, Phys. Rev. B {\bf 68},
  165344 (2003).  

\bibitem{Rashba}
E. I. Rashba, Sov. Phys. Solid State {\bf 2,} 1109 (1960).

\bibitem{Ando}
T. Ando,
Phys. Rev. B {\bf 40,} 5325 (1989).

\bibitem{Nitta}
J. Nitta, T. Akazaki, H. Takayanagi, and T. Enoki,
Phys. Rev. Lett. {\bf 78,} 1335 (1997).

\bibitem{Aleiner_PRL87_256801}
I. L. Aleiner and V. I. Fal'ko,
Phys. Rev. Lett. {\bf 87,} 256801 (2001).

\bibitem{Aleiner_PRL87_256801_Erratum}
I. L. Aleiner and V. I. Fal'ko,
Phys. Rev. Lett. {\bf 89,} 079902(E) (2002).

\bibitem{Oheun}
J. -I. Ohe, T. Ohtsuki and S. Murakami,
(unpublished).

\bibitem{Shinjo}
T. Shinjo, T. Okuno, R. Hassdorf, K. Shigeto and T. Ono,
Science {\bf 289,} 930 (2000).

\bibitem{PhysRevB.64.241201} 
T. Dietl, J. K\"onig and A. H. MacDonald,
Phys. Rev. B {\bf 64,} 241201(R) (2001).

\bibitem{Nazmul}
A. M. Nazmul, S. Sugahara and M. Tanaka,
Phys. Rev. B {\bf 67,} 241308(R) (2003).

\end{thebibliography}

\end{document}